\input{aipcheck}

\documentclass[sort&compress
    ,final            
  ]
  {aipproc}
\layoutstyle{8x11single}
\def\etal#1{ {\em et al.}}
\def\tit#1{}
\def\ii{{\rm i}}
\def\lan{\langle}
\def\ran{\rangle}
\def\llan{\left\langle}
\def\rran{\right\rangle}
\usepackage{amsmath,amssymb,latexsym}
\usepackage{graphicx,bbm}
\makeatletter
\def\underbracket{\@ifnextchar [ {\@underbracket} {\@underbracket [\@bracketheight]}}
\def\@underbracket[#1]{\@ifnextchar [ {\@under@bracket[#1]} {\@under@bracket[#1][0.4em]}}
\def\@under@bracket[#1][#2]#3{
           \mathop {\vtop {\m@th \ialign {##\crcr $\hfil \displaystyle {#3}\hfil $%
                              \crcr \noalign {\kern 3\p@ \nointerlineskip }\upbracketfill {#1}{#2}
                              \crcr \noalign {\kern 3\p@ }}}}\limits}
\def\upbracketfill#1#2{$\m@th \setbox \z@ \hbox {$\braceld$}
                  \edef\@bracketheight{\the\ht\z@}\bracketend{#1}{#2}
                  \leaders \vrule \@height #1 \@depth \z@ \hfill
                  \leaders \vrule \@height #1 \@depth \z@ \hfill \bracketend{#1}{#2}$}
\def\bracketend#1#2{\vrule height #2 width #1\relax}
\def\downbracketfill#1#2{$\m@th \setbox \z@ \hbox {$\braceld$}
                  \edef\@bracketheight{\the\ht\z@}\downbracketend{#1}{#2}
                  \leaders \vrule \@height #1 \@depth \z@ \hfill
                  \leaders \vrule \@height #1 \@depth \z@ \hfill
\downbracketend{#1}{#2}$}
\def\downbracketend#1#2{\vrule depth #2 width #1\relax}
\makeatother

\begin{document}

\title{Time series, correlation matrices and random matrix models}


\author{Vinayak}{
  address={Instituto de Ciencias F\' isicas, Universidad Nacional Aut\' onoma de M\' exico, C.P. 62210 Cuernavaca, M\' exico},email={vinayaksps2003@gmail.com}
}
\author{Thomas H. Seligman}{
  address={Instituto de Ciencias F\' isicas, Universidad Nacional Aut\' onoma de M\' exico, C.P. 62210 Cuernavaca, M\' exico}
  ,altaddress={Centro Internacional de Ciencias, C.P. 62210 Cuernavaca, M\' exico}, email={seligman@ce.fis.unam.mx} 
}

\maketitle
\section{Introduction}\label{intro}
In this set of five lectures the authors have presented techniques to analyze open classical and quantum systems using correlation matrices. For diverse reasons we shall see that random matrices play an important role to describe a null hypothesis or a minimum information hypothesis for the description of a quantum system or subsystem. In the former case various forms of correlation matrices of time series associated with the classical observables of some system. The fact that such series are necessarily finite, inevitably introduces noise and this finite time influence lead to a random or stochastic component in these time series. By consequence random correlation matrices have a random component, and corresponding ensembles are used. In the latter we use random matrices to describe high temperature environment or uncontrolled perturbations, ensembles of differing chaotic systems etc.

 The common theme of the lectures is thus the importance of random matrix theory in a wide range of fields in and around physics. The quantum aspects of these lectures were essentially covered in a course delivered by C. Pineda and one of the authors(THS) at the ELAF 2007. The lecture notes appeared in the corresponding proceedings \cite{Pinsel-AIPproc}. We will give a short overview in this introduction including references. Also we will highlight two new developments of applications of RMT in fidelity decay and decoherence that have occurred since, but we will not return to this subject in the successive sections.

For the treatment of fidelity decay \cite{Go-S-NJP-random, Ps-S-ecofid, Go-S-Review} and/or decoherence \cite{Go-S1, Go-S2, Pi-S-PRE, Go-S-NJP1} we assume that the couplings that destroy fidelity and cause decoherence are given as Gaussian random matrix ensembles \cite{Mehta, Brody81, GuhrGW98, Stoeckmann}. The same holds for the Hamiltonian of the environment. Quantities such as fidelity, purity, entropy, etc are calculated for members of the ensemble by performing partial traces and then averaged over the ensemble. The new developments not included are basically two: On one hand we presented an alternate Random Matrix procedure, in which we calculate the average density matrix in Ref. \cite{Go-S-NJP2}, and then calculate specific properties from that density matrix in  Refs. \cite{MOI-tesis, HEC-tesis}. The results we find, are very similar between the two approaches, but the meaning is rather different. The average destiny matrix is still a density matrix and thus describes an open system, whose specific properties we may test by measuring certain properties of a part of some entangled system or it may describe an ensemble as in the older work. Usually the latter is experimentally more accessible. Furthermore, we developed the random matrix theory of isospectral perturbations in Ref. \cite{Le-Gar-S} and proceeded to calculate the ensuing fidelity decay as suggested in a semiclassical investigation of the effect of isospectral perturbation \cite{Heller}. Unitary random matrix ensembles play an important role \cite{Le-Gar-S}, and we find a very particular case of the fidelity freeze predicted by Prosen and \v Znidari\v c \cite{freeze1,freeze2} with a random matrix version appearing in  Ref. \cite{freeze3}. Yet the original course is still essentially valid as an introduction. We shall therefore not return to this subject but rather concentrate in the present notes on the classical part of the course and how to use correlated stochastic time series to analyze real life problems. Because of the easy availability of financial data the more practical steps will rely on these, but we shall also use models from statistical physics and we may have in mind as widely different systems as chemical reactors, biological systems, medical applications or climate analysis.

We begin with a section on basic definitions of correlation matrices and a brief discussion of their application to financial markets in order to give some motivation. Next, we proceed to discuss the Wishart ensemble \cite{Wishart} which gives us a null hypothesis by showing the effects of random finite-time stochastic processes. A wide range of analytical results are available for this old problem, some of which we discuss in detail. We move on to a more recent subject, namely correlated random matrices \cite{vp2010} and we shall see, the effect of correlations on spectra. In the next section we will describe rather recent work  on a the Wishart model for nonsymmetric correlation matrices and study effects of correlations in this context \cite{vinasm}, which seem immediately useful for time delayed or time-lagged correlations. In the next section, we take up a very recent subject, namely the use of the power mapping, originally designed for noise suppression \cite{GuhrKabler,GuhrShafer2010}, as a sensitive detector of correlations \cite{VRT}. Finally, we give an outlook.

\section{Correlation matrices}\label{corrmat}
In order to give a hands on example of correlation matrices we introduce the concept using the example of financial time series. Suppose we have a stochastic system of $N$ real variables and we a record time series for each variable. For instance, let $S^{k}_{\tau}$ be the asset price for the $k$'th company at time $\tau$ from a market consisting of $N$ companies  or similarly the reading for the $k$'th electrode for a reading at time $\tau$ of an electroencephalogram
or electrocardiogram using $N$ electrodes. Then the covariance matrix, $\Sigma$, can be calculated as
\begin{equation}\label{covM}
\Sigma_{jk}= \langle(S^{j}_{\tau}-\mu_{j})(S^{k}_{\tau}-\mu_{k})\rangle_{T},
\end{equation}
where we have used $\langle\,.\,\rangle_{T}$ to represent averaging over the time for the time horizon $T$ and $\mu_{l}$ is the mean for the $l$'th variable. The correlation matrix, $\mathsf{C}$, is related with $\Sigma$ as
\begin{equation}\label{corrM}
{C}_{jk}=\frac{\Sigma_{jk}}{\sigma_{j}\sigma_{k}},
\end{equation}
where $\sigma_{l}$ is the standard deviation for the $l$'th variable. Note that $\mathsf{C}$ is a real symmetric matrix where $C_{jk}$ gives the cross-correlation between the $j$'th and the $k$'th variable.

In the financial markets \cite{Thomas2012,Finance1,Finance2,Finance3,Finance4}, however, it is more feasible to deal with the price change or {\it return},
over a time scale $\Delta \tau$. We define
\begin{equation}
R^{j}_{\tau}\equiv \ln\, S^{j}_{\tau+\Delta\tau}-\ln\,S^{j}_{\tau}.
\end{equation}
Since different stocks may have different standard deviations we normalize the time series as
\begin{equation}
A_{j\tau}=\frac{R^{j}_{\tau}-\mu_{j}}{\sigma_{j}},
\end{equation}
where $\mu_{l}$ and $\sigma_{l}$ are the mean and the standard deviation for the time series of return of the $l$'th return. Then the correlation matrix can be written as
\begin{equation}
\mathsf{C}=\mathsf{AA}^{t},
\end{equation}
where $\mathsf{A}^{t}$ is the transpose of the matrix $\mathsf{A}$.

A few remarks are immediate here. First we have $0\le|C_{jk}|\le1$ because of the normalization and the diagonal elements take the value $C_{jj}=1$. Next, because of the dyadic structure $\mathsf{AA}^{t}$ it is a positive semi-definite matrix meaning it has non-negative eigenvalues. Moreover, since $\mathsf{A}$ has rank $\{N,T\}_{min}$ the matrix $\mathsf{C}$ has the same rank. For $T \ge N$, the matrix is positive definite while for $T<N$ at least $N-T$ eigenvalues are zero. In the latter case $\mathsf{C}$ will be a positive semi-definite singular matrix. Note that the correlation matrix, as defined above, is an equal-time correlation matrix because we have considered values at the same time; see Eqs. (\ref{covM}) and (\ref{corrM}). Generalizing the correlation matrix in this sense, we define a time-lagged correlation matrix \cite{John,Potters:2005,Bouchaud:2009} which describes correlations among the variables at a time-lag $\Delta T$. In this case
\begin{equation}\label{asmcorr}
{C}_{jk}(\Delta T)=\langle A_{j,\tau}\,A_{k,\tau+\Delta T}\rangle_{T},
\end{equation}
which of course is not symmetric if $\Delta T\ne0$. In quantitative finance, both types of matrices are important; the equal time correlation matrices are mostly used for risk assessment purposes and the time-lagged matrices seem more relevant for forecasting models.

The {\it actual} correlations are defined for $T\rightarrow\infty$, which is much more theoretical concept. Finiteness of $T$ results in noise in the correlation matrix elements. There have been various methods \citet{other_tech1,other_tech2,GuhrKabler,GuhrShafer2010} to reduce noise from $\mathsf{C}$. Among these, noise reduction from the RMT approach gained much popularity \cite{gene, Seba, diverseAT,Sesmic}. RMT provides a general framework to see the generic impact of the noise in an empirical data sets. From RMT we use the Wishart models \cite{Wishart,Bouchaud:2009,vinasm}, which characterize a null hypothesis for correlation matrices. In these models all the variables are statistically independent and the time series are defined as white-noise time series.

\section{Wishart Models for symmetric correlation matrices}\label{symcorr}
\subsection{Wishart and Correlated Wishart Ensembles}
A Wishart matrix is defined as $\mathsf{C}=\mathsf{AA}^{t}/T$ where $\mathsf{A}$ is an $N\times T$ matrix, $\mathsf{A}^{t}$ is the transpose of $\mathsf{A}$ and entries of $\mathsf{A}$ are real independent Gaussian variables with zero mean and variance $\sigma^{2}$. In RMT the ensemble of Wishart matrices is studied in great detail and known as Wishart orthogonal ensemble (WOE) \cite{Mehta}. In the context of time series $\mathsf{C}$ may be interpreted as the covariance matrix calculated over stochastic time series of time horizon $T$ for $N$ statistically independent variables. This means on average $\mathsf{C}$ does not have cross-correlations. In the case of the actual cross-correlations one defines correlated Wishart orthogonal ensembles (CWOE) \cite{Wilks,Muirhead}. For instance, in order to take account of the correlations among the variables, we consider $\mathsf{C}=\xi^{1/2}\mathsf{BB}^{t}\xi^{1/2}$ where $\xi$ is a fixed positive definite matrix defining the actual correlations. For example, $\xi_{jj}=1$ represent the self-correlations. Matrix elements of $\mathsf{B}$ are independent Gaussian variables with zero mean and variance $\sigma^{2}$, just as the $A_{jk}$'s defined for WOE. It is not difficult to see how WOE is generalized in this model. For instance, $\mathsf{C}=\sigma^{2}\xi$ for $T\rightarrow\infty$. Equivalently, for finite $T$, we get $\overline{\mathsf{C}}=\sigma^{2}\xi$, where bar denotes the ensemble averaging. Finally, for $\xi=\mathbf{1}$, CWOE reduces to the WOE.
Using the Gaussian probability measure, the joint probability density (JPD) of the matrix elements of $\mathsf{B}$ can be written as
\begin{equation}\label{jpdajk}
P(\mathsf{B})\,d\mathsf{B}\propto \exp\left[-Tr\,\frac{\mathsf{BB}^{t}}{2\sigma^{2}}\right]d\mathsf{B},
\end{equation}
where $d\mathsf{B}$ is the infinitesimal volume in $N\times T$ matrix element space. For $T\ge N$, the JPD of the matrix elements of $\mathsf{C}$ is defined in $N(N+1)/2$-dimensional matrix element space. This JPD has been calculated by Wishart \cite{Wishart,Wilks,Muirhead}:
\begin{eqnarray}\label{jpdcjk}
P(\mathsf{C})\,d\mathsf{C}\propto (det\xi)^{-T/2}({det}\mathsf{C})^{[N(\kappa-1)-1]/2}\,
\exp\left[-\frac{T}{2\sigma^{2}}\,{Tr}\,\xi^{-1}\mathsf{C}\right]\,d\mathsf{C},
\end{eqnarray}
where $\kappa=T/N$ and $d\mathsf{C}=\prod_{j> k}^{N}dC_{jk}\prod_{j=1}^{N}dC_{jj}$ is the infinitesimal volume in matrix element space. Moreover, for $\xi=\mathbf{1}$, $C_{jk}$ can be described by a large $T$ expansion \cite{GuhrKabler}:
\begin{equation}\label{CvsT}
C_{jk}=\delta_{jk}+\sqrt{\frac{1+\delta_{jk}}{T}}\,a_{jk}+ \mathsf{O}\left(\frac{1}{T}\right), 
\end{equation}
where the $a_{jk}$ are independent Gaussian variables with zero mean and variance $\sigma^{2}$. Note that this expansion defines an ensemble similar to GOE except that here the ensemble average of each diagonal term is $1$.

\subsection{Spectral Statistics of WOE}
For WOE, the JPD of the eigenvalues of $\mathsf{C}$, $\lambda_{j}$, for $1\le j\le N$, can
be obtained using techniques developed for the Gaussian ensembles \cite{Mehta}.
For instance, we use $\xi=1$ in JPD (\ref{jpdcjk}) and transform it to eigenvalue-eigenvector
space and integrate out the eigenvectors. Then we get the JPD of eigenvalues:
\begin{eqnarray}\label{jpdigvl}
P(\lambda_{1},...,\lambda_{N})\propto \prod_{j=1}^{N}w(\lambda_{j})\prod_{j>k}^{N}|\lambda_{j}-\lambda_{k}|.
\end{eqnarray}
Here $w(\lambda)=\lambda^{[N(\kappa-1)-1]/2}\exp[-N\kappa \lambda/2\sigma^{2}]$ is
the weight function of the associated Laguerre polynomials, hence WOE is often referred
to as Laguerre orthogonal ensemble in the literature \cite{Mehta,Brody81,GuhrGW98,BenRMP97}.
The Vandermonde determinant in Eq. (\ref{jpdigvl}) comes from the Jacobian of the
transformation from matrix-element space to eigenvalue-eigenvector space. In analogy to
Gaussian orthogonal ensemble (GOE), where all the spectral correlations are known in
terms of Hermite polynomials, for WOE all the spectral correlations are known in terms
of Laguerre polynomials. For GOE, the spectral density converges to Wigner's semi-circle
at large matrix dimension. Similarly, for WOE, for large $N$ and $T$ with finite $\kappa$
we get the  Mar\'{c}enko Pastur density \cite{marchenko}:
\begin{equation}
\label{denmp}
\overline{\rho}(\lambda)=\kappa\frac{\sqrt{(\lambda_{+}-\lambda)(\lambda-\lambda_{-})}}{2\pi \sigma^{2}\lambda},
\end{equation}
where $\lambda_{\pm}=\sigma^{2}(\kappa^{-1/2}\pm 1)^2$ are the end points of the density. 
Note that for positive semi-definite matrices, i.e., for $\kappa<1$, the density $\overline{\rho}(\lambda)$
in the above equation is normalized to $\kappa$ and not to $1$. Therefore, taking into account the $(N-T)$ zeros, for $\kappa\le1$ we write
\begin{equation}
\label{denmpkle1}
\overline{\rho}(\lambda)=\kappa\frac{\sqrt{(\lambda_{+}-\lambda)(\lambda-\lambda_{-})}}{2\pi \sigma^{2}\lambda}+(1-\kappa)\delta(\lambda).
\end{equation}

Higher order spectral correlations are described by the $n$-point spectral correlation function where $n\ge2$. The $n$-point spectral correlation function
describes statistics of $n<N$ eigenvalues of the spectra irrespective of the remaining $N-n$ eigenvalues. For example, the spectral density is the
one-point function. The average nearest neighbor spacing distribution is, maybe, the most popular manifestation of the spectral fluctuation properties. Yet it involves general $n$-point functions, while two-point functions e.g. form factors or number variances are easier to calculate. Intuitively, it may be understood from the expansion (\ref{CvsT}) why WOE and GOE must have the same spectral fluctuations. The number of the off-diagonal terms of a matrix of dimension $N$ is of order $N^{2}$ while the number of diagonal terms is $N$ and therefore the former dominates except in the case of the one-point function where diagonal terms enter on a different footing than the off-diagonal ones. For a rigorous proof we refer to \cite{APSG} where skew-orthogonal polynomials have been used to calculate the $n$-point correlation functions. This agreement between WOE and GOE supports the universality of spectral fluctuations which is a main cause behind the success of RMT in a wide variety of fields.

\subsection{Pastur Density and Universal Fluctuations of CWOE}

In the presence of the actual cross-correlations, i.e., for $\xi\ne \mathbf{1}$, integration over the group of orthogonal matrices, which we need to obtain the JPD of the eigenvalues, has not been possible yet. Because of this bottleneck analytic results for CWOE are limited. For instance, the Pastur self-consistent equation \cite{Pastur} which describes the spectral density for large matrices is known for CWOE but for fluctuations only the asymptotic result for the two-point function is known \cite{vp2010}. The latter describes spectral regions having universal fluctuations.

The Pastur equation can be obtained by using the binary correlation method. The binary correlation method was developed by French and his collaborators \cite{Brody81,ap81}. In the binary correlation method we deal with the resolvent or the Stieltjes transform of the density, defined as,
\begin{equation}
\overline{G}(z)=\frac{1}{N} {Tr} \overline{(z-\mathsf{C})^{-1}}.
\end{equation}
The spectral density $\overline{\rho}(\lambda)$ can be determined uniquely via the inverse transformation:
\begin{equation}
\label{denCWE}
\overline{\rho}(\lambda)=\mp \frac{1}{\pi}\,\Im\,\overline{G}(\lambda\pm\ii\epsilon),
\end{equation}
where $\epsilon>0$ is infinitesimal. The ensemble averaged resolvent for the CWOE is described by a self-consistent Pastur equation,
\begin{equation}\label{ResMp}
\overline{G}(z)=\frac{1}{N}{Tr} \left(z-\frac{\sigma^{2}}{\kappa}\left(\kappa-1+z\overline{G}(z)\right)\xi\right)^{-1}.
\end{equation}
We refer to the first paert of the Appendix for a step by step derivation of this result. Ironically, this result was derived by Mar\'cenko and Pastur \cite{marchenko} in $1967$ but it remained almost unnoticed against the famous Mar\'cenko Pastur law which is actually a consequence of this result for $\xi=\mathbf{1}$. The same result has been obtained by others \cite{Silverstien,SenM,Burda:2005,vp2010} using different techniques. For $\xi=\mathbf{1}$, the Pastur equation (\ref{ResMp}) results a quadratic equation in $\overline{G}(z)$. Note that while solving the quadratic equation in the complex plane we encounter branch-point singularities. However, it turns out that to obtain a positive density we must admit the branch, where the solution $\overline{G}(z)\sim z^{-1}$ for large $z$. The inverse transform (\ref{denCWE}) then yields the Mar\'cenko Pastur law (\ref{denmp}).

For a non-trivial spectrum of $\xi$, Eq. (\ref{ResMp}) has to be solved numerically. In Ref. \cite{vp2010} we have developed a numerical technique to solve this equation. We basically use Newton's method to solve the equation $F(G,z)-G(z)=0$, where $F(G,z)$ is the right hand side of the Eq. (\ref{ResMp}) and $z=x+\ii\epsilon$ is fixed. We start with an initial guess for $G$. When the solution converges to a desired precision (machine precision) we stop the iterations and use the solution as a guess for $G$ in the neighborhood of $z$ on the real axis.

\begin{figure}
        \centering
               \includegraphics [width=0.5\textwidth]{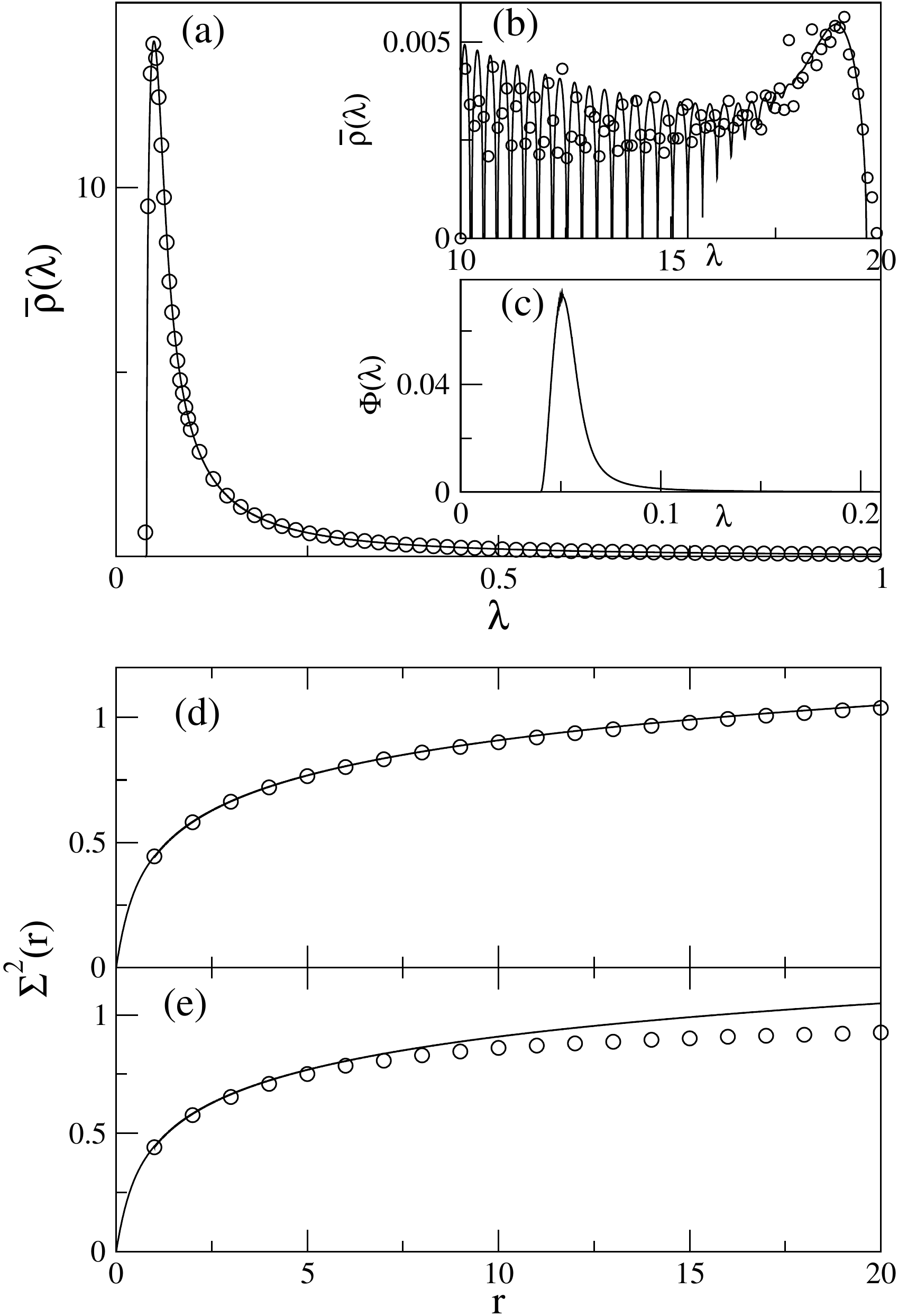}
                \caption{Density and fluctuation analysis for CWOE where $\xi_{jk}=c^{|j-k|}$,
$c=0.9$. The solid line in (a) is the theoretical density obtained by solving Eq. (\ref{ResMp})
numerically. In (b) we show the tail of the same density. Open circles in all figures represent
results obtained from the diagonalization of $\mathsf{C}$. In (c) we plot the comparison
function defined in (\ref{comp}). In (d) and (e) we analyze fluctuations about the density.
We consider the number variance to compare with GOE result shown in both figures by solid lines.
Spectral regions in (d) and (e) are respectively $0.045<\lambda<0.15$ and $0.15<\lambda<0.3$.}
\label{CWEtp}
\end{figure}

Using the binary correlation method we can also calculate the leading order result for the two-point Stieltjes transform, defined as
\begin{eqnarray}\label{SG}
S^{G}(z_{1},z_{2})=\overline{G(z_{1})G(z_{2})}-\overline{G(z_{1})}\,\overline{G(z_{2})}.
\end{eqnarray}
If we can calculate $S^{G}(z_{1},z_{2})$ then we can also determine the two-point function, $S^{\rho}(\lambda_{1},\lambda_{2})$. The two-point function is defined as
\begin{eqnarray}\label{Srho}
S^{\rho}(\lambda_{1},\lambda_{2})=\overline{\rho(\lambda_{1})\rho(\lambda_{2})}-\overline{\rho(\lambda_{1})}\,\overline{\rho(\lambda_{2})}.
\end{eqnarray}
It is related with $S^{G}(z_{1},z_{2})$ via
\begin{eqnarray}
S^{\rho}(\lambda_{1},\lambda_{2})
&=&
-\frac{1}{4\pi^{2}}\Big[
S^{G}(z_{1}^{+},z_{2}^{+})+S^{G}(z_{1}^{-},z_{2}^{-})
-S^{G}(z_{1}^{+},z_{2}^{-})-S^{G}(z_{1}^{-},z_{2}^{+})
\Big].
\end{eqnarray}
Here $z^{\pm}=\lambda\pm \ii \epsilon$. As for the density, here as well, we compute only the leading order term for $S^{G}(z_{1},z_{2})$. This term is of order ${O}(N^{-2})$ because the terms of ${O}(1)$ and of ${O}(N^{-1})$ cancel exactly; see Ref. \cite{vp2010} for the details. We need the two-point function to derive the number variance, $\Sigma^{2}(r)$, which is usually used as a measure for the long-range spectral correlations. It is defined as the variance of the number of eigenvalues in an interval of length $r$ which on average has $r$ eigenvalues \cite{Brody81}. There exists a one to one correspondence between the number variance and the rescaled two-point correlation function \cite{Verbaarschot}, which is more convenient to see short range correlations. As mentioned above, we need all $n$-point correlations to obtain the nearest neighbor spacing distribution, yet at short range this very popular measure coincides with the two-point function; the differences arise, because the two-point function allows eigenvalues between the two values considered, which is very unlikely for universal fluctuations at short distances due to the linear repulsion of eigenvalues in the orthogonal ensembles (and even stronger repulsion in the other universal cases not considered here; for random eigenvalues the situation is more subtle). Note that in definitions of most fluctuation measure the measures the spectra is expressed in terms of average nearest neighbor spacings, a procedure which is known as the {\it unfolding}. As we have mentioned above that because of certain limitations, for CWOE, such a detailed result has not been possible yet.

For CWOE, after some non-trivial algebra, we find that
\begin{equation}\label{unitp}
\frac{S^{\rho}(\lambda_{1},\lambda_{2})}{\overline{\rho}(\lambda_{1})\,\overline{\rho}(\lambda_{2})}
=
-\frac{1}{\pi^{2} r^{2}},
\end{equation}
provided
\begin{equation}\label{crit}
2\pi\overline{\rho}^{2}N>> 
\left|
\left(
\frac{1-\kappa}{\lambda^{2}}-\frac{\partial\overline{ \mathsf{P}}}{\partial \lambda}
\right)r
\right|.
\end{equation}
Here $\lambda=(\lambda_{1}+\lambda_{2})/2$ is the mean position, $r=|\lambda_{1}-\lambda_{2}|N\rho(\lambda)$
is the spectral correlation length, and
\begin{equation}
\overline{\mathsf{P}}(\lambda)=\Re\,\overline{G}(\lambda\pm \ii \epsilon),
\end{equation}
is the principal value integral of the density. It is the same result that we obtain
for the GOE \cite{Brody81} and the WOE \cite{vinayak}, for $r>1$. This result
is in agreement with the leading order non-periodic term of the asymptotic
expansion of the GOE result \cite{Mehta}. Hence, Eq. (\ref{crit}) describes
the spectral regions where
the spectral fluctuations are universal. If the inequality does not hold in a
spectral region, for large $r$, we expect deviations in $\Sigma^{2}(r)$ from
the universal prediction. For example, let $\xi_{jk}=c^{|j-k|}$ where $c=0.9$.
In Fig. \ref{CWEtp} we show $\overline{\rho}(\lambda)$ and $\Sigma^{2}(r)$.
In Fig. \ref{CWEtp}(a) we compare the density obtained from the Monte Carlo
simulations of $\mathsf{C}$ with the theoretical density obtained from Eqs.
(\ref{ResMp}) and (\ref{denCWE}). In Fig. \ref{CWEtp}(b) we show long tail
of the density. In Fig. \ref{CWEtp}(c) we plot a comparison function
\begin{equation}\label{comp}
\Phi(\lambda)=
\frac{2\pi\overline{\rho}^{2}}{(1-\kappa)\lambda^{-2}-\partial\overline{ \mathsf{P}}/\partial \lambda}.
\end{equation}
The inequality predicts universal fluctuations when $\Phi>>N^{-1}$.
Therefore in the region $0.045<\lambda<0.15$ where this function has
a peak we indeed find universal fluctuations; see Fig. \ref{CWEtp}(d).
On the other hand in the tail, $0.15<\lambda<0.3$, we see deviations
in Fig. \ref{CWEtp}(e). This is consistent with the prediction of
Fig. \ref{CWEtp}(b). Indeed, as predicted by $\Phi(\lambda)$, the deviation
becomes stronger the farther we reach into the tail. Short range fluctuations,
however, are consistent with universal predictions in both the regions.

Even more interesting is the case of separation of few eigenvalues from the bulk
of the density \cite{Finance1, Finance2,Finance3, Finance4, Baik:2005, Forrester,vp2010,vmarko}.
These eigenvalues often show collective behavior, therefore referred to
as the collective modes. The mean position of these collective modes can be
calculated by using the Pastur equation while for the variance we need the
two-point function \cite{bulk}. Asymptotic results for the collective modes are
also given in \cite{vp2010}. The collective modes are of importance in the
correlation matrix analysis of several systems \cite{Finance1, Finance2,Finance3, 
Finance4, Sesmic, vmarko}.

\section{Wishart Model for Non-symmetric Correlation Matrices}\label{asmcrr}

We have explained the section about correlation matrices that the time-lagged correlation matrices are
not symmetric. The random matrix model which we may define in this case is comprised
of two statistically equivalent but different matrices $\mathsf{A}$ and $\mathsf{B}$
and is defined as $\mathsf{AB}^{t}$. Like the equal-time correlation matrices, for
time-lagged matrices, such model can be used as a null hypothesis \cite{Bouchaud:2007}
where we consider statistical independence of $\mathsf{A}$ and $\mathsf{B}$. As a
matter of fact this model can be used for a more general case where the correlation
matrix describes statistics between two different statistical system
\cite{Muller,Stanley:2011,Livan:2012}. For square matrices, statistics of the
eigenvalues define a reference against which the correlation must be viewed.
Several results are known in this case \cite{Burda:2010,Akeman:2010,BielyandThurner}.
For rectangular matrices, we study statistics of singular values or equivalently a
Wishart model defined as $\mathbf{C}=\mathsf{AB}^{t}\mathsf{BA}^{t}$ \cite{sing}.
Using the CWOE approach we generalize this model to the case where $\mathsf{A}$
and $\mathsf{B}$ have correlations, implying that the ensemble
average $\overline{\mathsf{AB}^{t}}\ne 0$.
\subsection{Generalities}
Let $\mathcal{C}=\mathcal{WW}^{t}$ where
\begin{eqnarray}
\mathcal{W}=
\left(
\begin{matrix}
\mathcal{A} \\
\mathcal{B} 
\end{matrix}
\right),
\end{eqnarray}
Here matrices $\mathcal{A}$ and $\mathcal{B}$ are of dimensions
$N\times T$ and $M\times T$, respectively. Then the matrix $\mathcal{C}$
is a partitioned matrix, defined in terms of $\mathcal{A}$ and $\mathcal{B}$, as
\begin{equation}\label{CWEC}
\mathcal{C}=\dfrac{1}{T} 
\left(
\begin{matrix}
\mathcal{AA}^{t} & \mathcal{AB}^{t}\\
\mathcal{BA}^{t} & \mathcal{BB}^{t}.
\end{matrix}
\right),
\end{equation}
Here the diagonal blocks, viz., $\mathcal{AA}^{t}$ and $\mathcal{BB}^{t}$, and
the off-diagonal blocks, viz., $\mathcal{AB}^{t}$ and $\mathcal{BA}^{t}$, are
respectively $N\times N$, $M\times M$, $N\times M$ and $M\times N$ dimensional.
Therefore $\xi$ is also partitioned:
\begin{equation}\label{CWEXI}
\xi=
\left( 
\begin{matrix}
\xi_{\text{AA}} & \xi_{\text{AB}}\\
\xi_{\text{BA}} & \xi_{\text{BB}}
\end{matrix}
\right),
\end{equation}
where the diagonal blocks $\xi_{\text{AA}}$ and $\xi_{\text{BB}}$ account for the
correlations among the variables of $\mathcal{A}$ and of $\mathcal{B}$,
respectively. The off-diagonal blocks, e.g., $\overline{\mathcal{AB}^{t}}/T=\xi_{\text{AB}}$,
account for the correlations of $\mathcal{A}$ and $\mathcal{B}$. By construction $\xi_{\text{BA}}=[\xi_{\text{AB}}]^{t}$. Without loss of generality we consider $M\ge N$ and $T\ge M$ .

We consider the case where $\xi_{\text{AB}}\ne 0$ and wish to compare the spectral
density with that of the null hypothesis, i.e., when $\xi_{\text{AA}}=\mathbf{1}_{N\times N}$, $\xi_{\text{BB}}=\mathbf{1}_{M\times M}$ and $\xi_{\text{AB}}=0$. It is therefore important to remove cross-correlations among the variables of individual matrices because only then will the diagonal blocks of (\ref{CWEC})
yield identity matrices upon ensemble averaging.
Thus we introduce decorrelated matrices \cite{vp2010} defined as
\begin{eqnarray}\label{decorr}
\mathsf{A}&=&\xi_{\text{AA}}^{-1/2}\mathcal{A},
\nonumber\\
\mathsf{B}&=&\xi_{\text{BB}}^{-1/2}\mathcal{B}.
\end{eqnarray}
Note that we still have $\overline{\mathsf{AB}^{t}}/T=\eta$ where
$\eta=\xi_{\text{AA}}^{-1/2}\,\xi_{\text{AB}}\,\xi_{\text{BB}}^{-1/2}$
and the null hypothesis is characterized by $\eta=0$. This case has been
studied by several authors \cite{Muller,Bouchaud:2007, Burda:2010} but a study incorporating the actual correlations in theory has been reported recently \cite{vinasm}.
In this study we consider $\eta\ne0$ and calculate the ensemble averaged spectral density of $N\times N$ symmetric matrices $\mathbf{C}$, defined as
\begin{equation}\label{C}
\mathbf{C}=\frac{\mathsf{AB}^{t}\mathsf{BA}^{t}}{T^{2}}.
\end{equation}
We further define the ratios, $\kappa_{N}=N/T$ and $\kappa_{M}=M/T$.

A few remarks are immediate. At first we note that the ensemble average yields,
$\overline{\mathbf{C}}=\kappa_{M}\textbf{1}_{N\times N}+\eta\eta^{t}$.
Next, for $T\rightarrow\infty$ and $N,M$ finite, since $\mathcal{C}=\xi$, $\mathbf{C}=\eta\eta^{t}$.  We finally define a symmetric matrix $\zeta$, as
\begin{equation}
\zeta=\eta\eta^{t}.
\end{equation}
However, in the following we consider a large $N$ limit where $N/T$ and $M/T$ are finite so that
$\mathbf{C}$ will never be deterministic. We consider only those cases where the spectrum of
$\zeta$ does not exceed $N$. This is always valid for our model because the positive
definiteness of $\xi$ ensures an upper bound $1$ for eigenvalues of $\zeta$. Interested
readers can find proof of the last remark in Ref. \cite{vinasm}.

\begin{figure}
        \centering
               \includegraphics [width=0.4\textwidth]{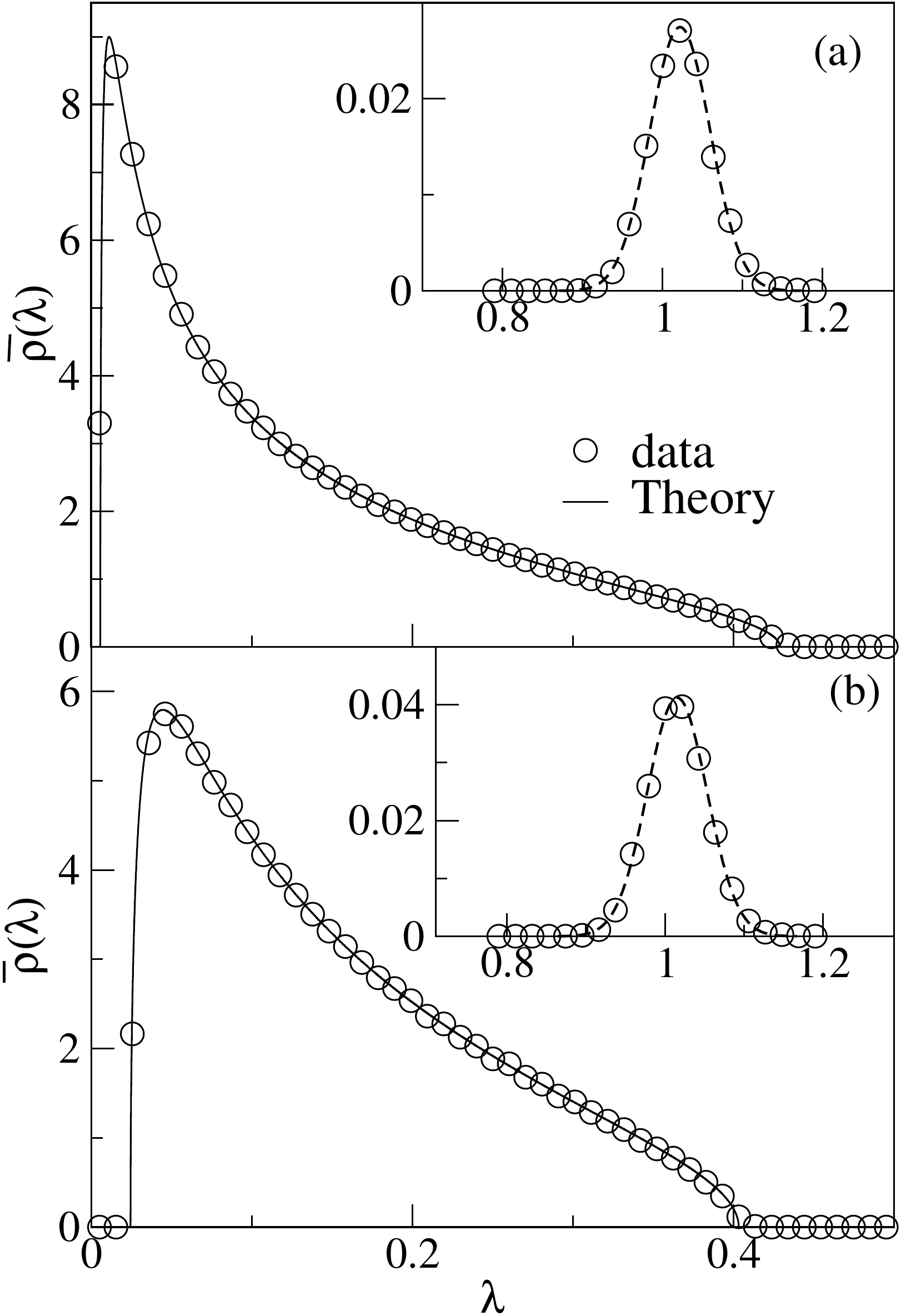}
                \caption{Spectral density, $\overline{\rho}(\lambda)$, where $[\xi_{\text{AB}}]_{jr}=c$ for $1\le j\le N$ and $1\le r\le M$, $c=0.8$, and correlation coefficients of the equal-cross-correlation matrices describing the diagonal blocks are $a=b=0.5$. Symbols in the figure represent Monte Carlo simulations and solid lines are the theory obtained from the numerical solution of Eq. (\ref{FinRes}). In (a) we show results for $N=384$ and in (b) we show results for $N=256$. Dimension of the full matrix in both the figures is $1024$ and $T=5120$. In insets we show distribution of the separated eigenvalues where we have considered an ensemble of $10000$ matrices. Dashed lines in insets represent Gaussian distributions where the means and the variances have been calculated numerically.}
\label{deneqc}
\end{figure}
\subsection{The Spectral Density}
Like in the CWOE case, here also the spectral density is described by a Pastur self-consistent equation. Using the binary correlation method here as well, we derive the Pastur equation for the spectral density. This equation may be written compactly as
\begin{equation}\label{FinRes}
\overline{G}(z)=
\llan 
\left(z-\zeta\overline{Y}_{1}(z,\overline{G}(z))-\overline{Y}_{2}(z,\overline{G}(z))\right)^{-1}
\rran,
\end{equation}
where
\begin{eqnarray}\label{Y1}
\overline{Y}_{1}(z,\overline{G}(z))&=& \frac{\left[1+\kappa_{N}\left(z\,\overline{G}(z)-1\right) \right]^{2}}
{1-\frac{\kappa_{N}\overline{G}(z)\left[1+\kappa_{N}\left(z\,\overline{G}(z)-1\right) \right]}{1-\kappa_{N}\overline{g}(z,\overline{G}(z))}},
\\
\label{Y2}
\overline{Y}_{2}(z,\overline{G}(z))&=&
\frac{\overline{Y}(z,\overline{G}(z))}{1-\kappa_{N}\overline{g}(z,\overline{G}(z))},
\\
\label{Y}
\overline{Y}(z,\overline{G}(z))&=&
\kappa_{M}+\kappa_{N}\left(z\,\overline{G}(z)-1\right)
\left[1+\kappa_{M}+\kappa_{N}\left(z\,\overline{G}(z)-1\right)
\right],
\end{eqnarray}
and
\begin{equation}\label{gY2}
\overline{g}(z,\overline{G}(z))=\frac{[z-\overline{Y}_{2}(z,\overline{G}(z))]\overline{G}(z)-1}{1+\kappa_{N}\left(z\,\overline{G}(z)-1\right)}.
\end{equation}
Eq. (\ref{FinRes}) together with definitions (\ref{Y1}-\ref{gY2}) is the Pastur equation which describes the spectral density. For the uncorrelated case, i.e., for $\zeta=0$, Eq. (\ref{FinRes}) reduces to a cubic equation confirming thereby the result obtained in Ref. \cite{Burda:2010}.

\subsection{Numerical examples}
We demonstrate our result for two different correlation matrices. In the first example we consider $\xi_{\text{AB}}$ to be a rank one matrix, e.g., $[\xi_{\text{AB}}]_{jr}=c$ for every integer $1\le j\le N$ and $1\le r\le M$. In the second example we consider $[\xi_{\text{AB}}]_{jr}=c\,\delta_{jr}+(1-\delta_{jr})c^{|j-r|}$. For simplicity, we consider that the diagonal blocks are chosen as the equal-cross-correlation matrix model, e.g., $[\xi_{\text{AA}}]_{jk}=\delta_{jk}+(1-\delta_{jk})a$, for $1\le j,k\le N$, and the same for $\xi_{\text{BB}}$ where the correlation coefficient is $b$. In both examples we consider $0<a,b,c<1$. This choice is necessary but not sufficient for the positive definiteness of $\xi$ \cite{vinasm}.

To check our theoretical result (\ref{FinRes}) with numerics we begin by simulating $\mathcal{C}$. Next, we identify the off-diagonal block $\{\text{AB}\}$ in $\mathcal{C}$. Finally, we use the transformation $\xi_{\text{AA}}^{-1/2}\mathcal{AB}^{t}\xi_{\text{BB}}^{-1/2}$ to obtain $\mathsf{AB}^{t}$ which we need to calculate $\mathbf{C}$. In numerical simulations we fix $N+M=1024$, $T=5(N+M)$ and consider two values of $N$, viz., $N=256$ and $384$. To compare numerics with theory we consider an ensemble of size $1000$ of matrices $\mathbf{C}$.
\begin{figure}[!t]
        \centering
               \includegraphics [width=0.4\textwidth]{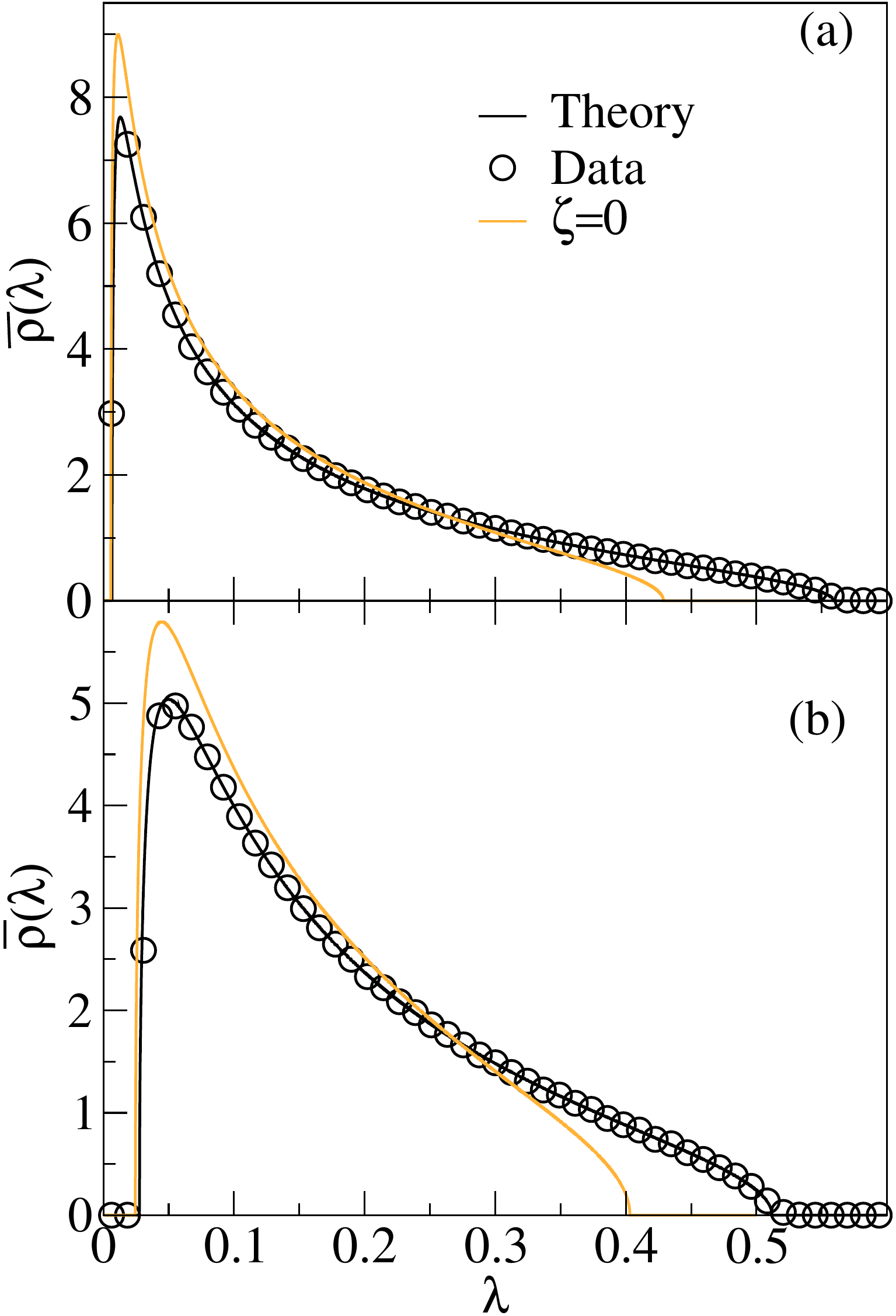}
                \caption{(Color online) Spectral density, $\overline{\rho}(\lambda)$, for the second example where $[\xi_{\text{AB}}]_{jr}=c\,\delta_{jr}+(1-\delta_{jr})c^{|j-r|}$, for $1\le j\le N$ and $1\le r\le M$, $c=0.05$ and the correlation coefficients which describe the diagonal blocks are $a=b=0.5$. With an outlay similar to Fig. \ref{deneqc} we compare our theory with numerics. Color lines in this figure represent the uncorrelated case.}
\label{dencomp}
\end{figure}

In our first example $\zeta$ has only one non-zero eigenvalue. For this spectrum our theory (\ref{FinRes}) yields the density for the bulk of the spectra. It suggests that the bulk should be described by the density of the uncorrelated case. We verify this with numerics in Fig. {\ref{deneqc}}, where $a=b=0.9$ and $c=0.8$. However, like in the equal-cross-correlation matrix model of CWOE we obtain one eigenvalue separated from the bulk \cite{bulk}; see Eq. (\ref{eqdeneqc}) for comparison. Interestingly, here the bulk remains invariant with correlations as opposed to the CWOE case where it changes with the correlations. Moreover, the distribution of the separated eigenvalues is well described by a Gaussian distribution \cite{dbulk} as shown in insets of Figs. \ref{deneqc}(a) and \ref{deneqc}(b).

Our second example corresponds to a non-trivial spectrum of $\zeta$. We consider the correlation matrix $\xi$ as explained above with parameters  $a=b=0.5$ and $c=0.05$. Note that the off-diagonal blocks have small contributions to the largest eigenvalues of $\xi$ and therefore are difficult to  trace in the analysis of separated eigenvalues of the corresponding CWOE. In Fig. \ref{dencomp} we compare our theory with numerics, for $N=384$ in Fig. \ref{dencomp}(a) and for $N=256$ in Fig. \ref{dencomp}(b). As shown in the figure, even small correlations in $\xi_{\text{AB}}$ cause notable changes in the density which are described well by the theory.

\section{Singular Correlation Matrices and The Power mapping Method}\label{PM}

Prime examples of CWOE applications can be seen in financial time series \cite{Finance4, Bouchaud:2009}. Applications of CWOE usually imply stationarity of the time series after eliminating some well known trends. We may well have a much larger number of time series than the number of time steps over which the time series can reasonably be considered as stationary \cite{Thomas2012}. This situation leads to correlation matrices that are highly singular. We shall use the power map \cite{GuhrKabler,GuhrShafer2010} to remove the degeneracy of zero eigenvalues, but we first give a brief view, how effective this method can be for noise reduction in singular correlation matrices.
\subsection{Noise Reduction By the Power Map}
\begin{figure}
        \centering
               \includegraphics [width=0.6\textwidth]{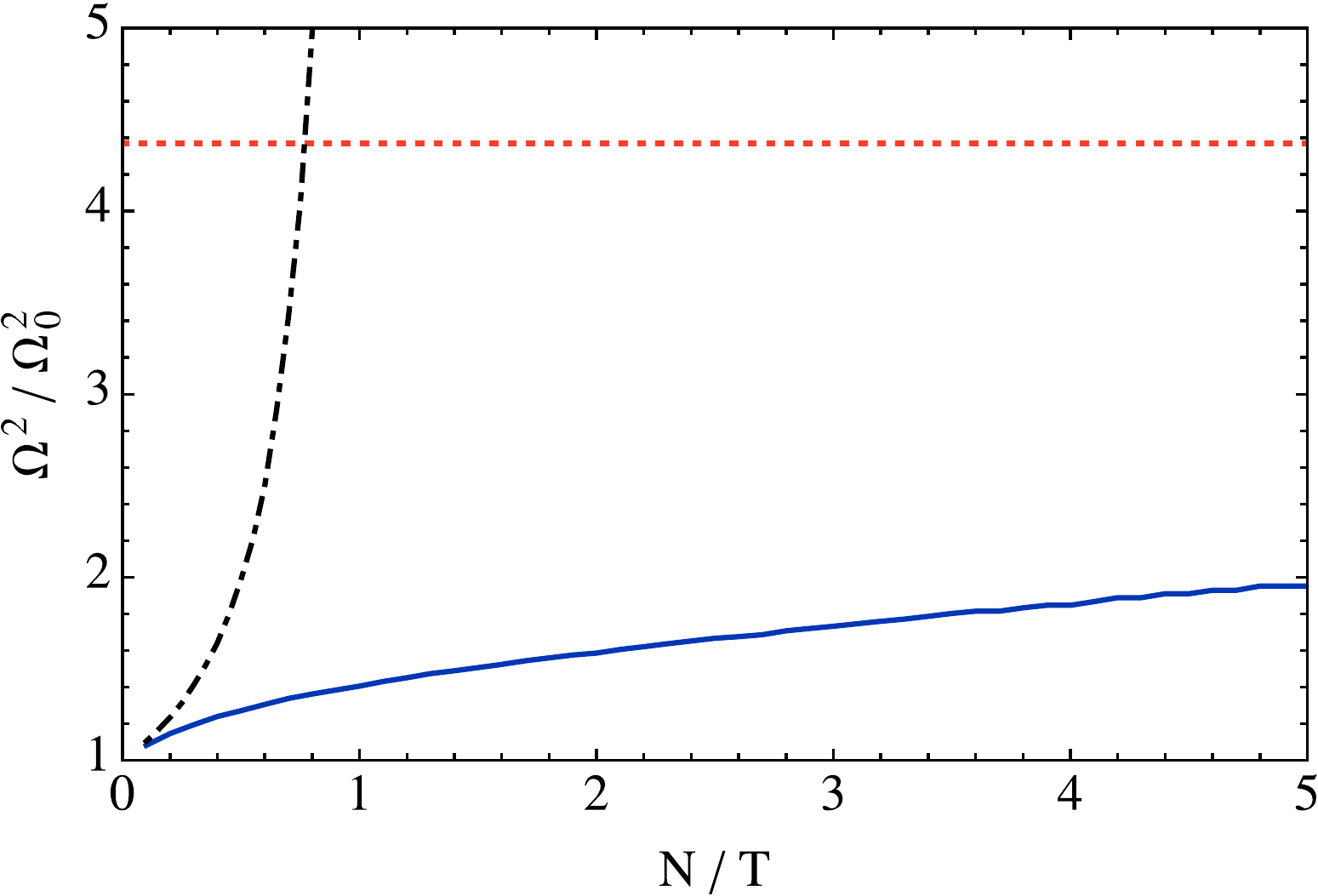}
                \caption{Portfolio variance $\Omega^2$ normalized to the minimal portfolio variance $\Omega_0^2$. The length $T$ of the time series was varied, the number of companies was fixed at $N=100$. Results are shown for sample correlations without noise reduction (black dashed-dotted line) and
for power-mapped correlations (blue solid line).The red dashed line at $\Omega^2/\Omega_0^2=4.37$ corresponds to a homogeneous portfolio.}
\label{pmbeyond}
\end{figure}
As the power map has mainly been discussed for {regular} or full rank correlation matrices we start giving a brief discussion of recently published \cite{VRT} results for singular correlation matrices from the realm of econophysics. In this section we shall use standard notations of econophysics as summarized in the second part of the Appendix. Following Markowitz~\cite{markowitz59}, we consider a portfolio of $N$ stocks and wish to calculate the portfolio weights $w_\mathrm{opt}$ which minimize the portfolio variance
\begin{equation}
\Omega^2=w_\mathrm{opt}^t \Sigma w_\mathrm{opt} \ .
\end{equation}
In a model setting, we can calculate the minimal variance portfolio $\Omega_0^2$ using the model covariance matrix $\Sigma_0$. In practice, however, the covariance matrix has to be estimated using historical data of finite length $T$. The shorter the length $T$ of the time series, the noisier is the covariance estimation. Using noisy covariance matrices for portfolio optimization leads to very bad results, see Fig.~\ref{pmbeyond}. In this case, the portfolio variance $\Omega^2$ increases as $(1-N/T)^{-1}$, in accordance with the literature~\cite{pafka03}. Clearly, it is necessary to improve the estimation of the covariance matrix to obtain better results. The variances of the single stocks can be estimated rather well on short time horizons due to a slowly decaying autocorrelation. The noise in the correlation coefficients $C_{kl}$ can be reduced effectively using the power map. This map operates directly on the correlation matrix elements by the simple means of elevating their absolute value to a power greater than one while conserving their phase
\begin{equation}\label{pmap}
C^{(q)}_{kl}={\rm sgn}\,C_{kl} \, \big| C_{kl} \big|^q \ .
\end{equation}
The results are presented in Fig.~\ref{pmbeyond}. The power map yields portfolio variances which are well below the homogeneous portfolio with all weights equal to $1/N$, even for $N>T$ where the correlation matrix becomes singular. The values for $q$ used in this study range from 1.1 to 2.4. For details of the simulation we refer to the second part of the Appendix.

\subsection{Statistics of the Emerging Spectra}

The non-linearity of the power map lifts the degeneracy of eigenvalues at zero. The thus emerging spectrum gives us a handle to get more information from the eigenvalues without looking at the entire correlation matrix as in \cite{Thomas2012}. The emerging spectra, we wish to study, may be observed even for $q$ very near to identity. Although the non-linearity of the power map makes it difficult, some analytic results for emerging spectra can be calculated for WOE and for a special case of CWOE where $\xi_{jk}=\delta_{jk}+(1-\delta_{jk})c$ and $0<|c|<1$. This model is referred to as the equal cross-correlation matrix model and often is important in applications.

Before developing an analytic approach, we mention a few important spectral properties of $\mathsf{C}^{(q)}$ as observed in simple numerical simulations for WOE. First we note that $\mathsf{C}^{(q)}$ is always real symmetric; thus it has real eigenvalues. However, for $q\ne1 $, it may have negative eigenvalues specially when $T$ is much smaller than $N$. The density function of the eigenvalues of $\mathsf{C}^{(q)}$, appears on two well separated supports. The first one is close to zero while the other is close to the support defined by the Mar\'{c}enko-Pastur density (\ref{denmp}). The former results from the breaking of degeneracy of the zero eigenvalues of $\mathsf{C}$ while the latter is due to small corrections to the original non-zero spectrum. As we increase the power to values usually used for noise reduction, the two supports begin to overlap.

\begin{figure}
        \centering
               \includegraphics [width=0.6\textwidth]{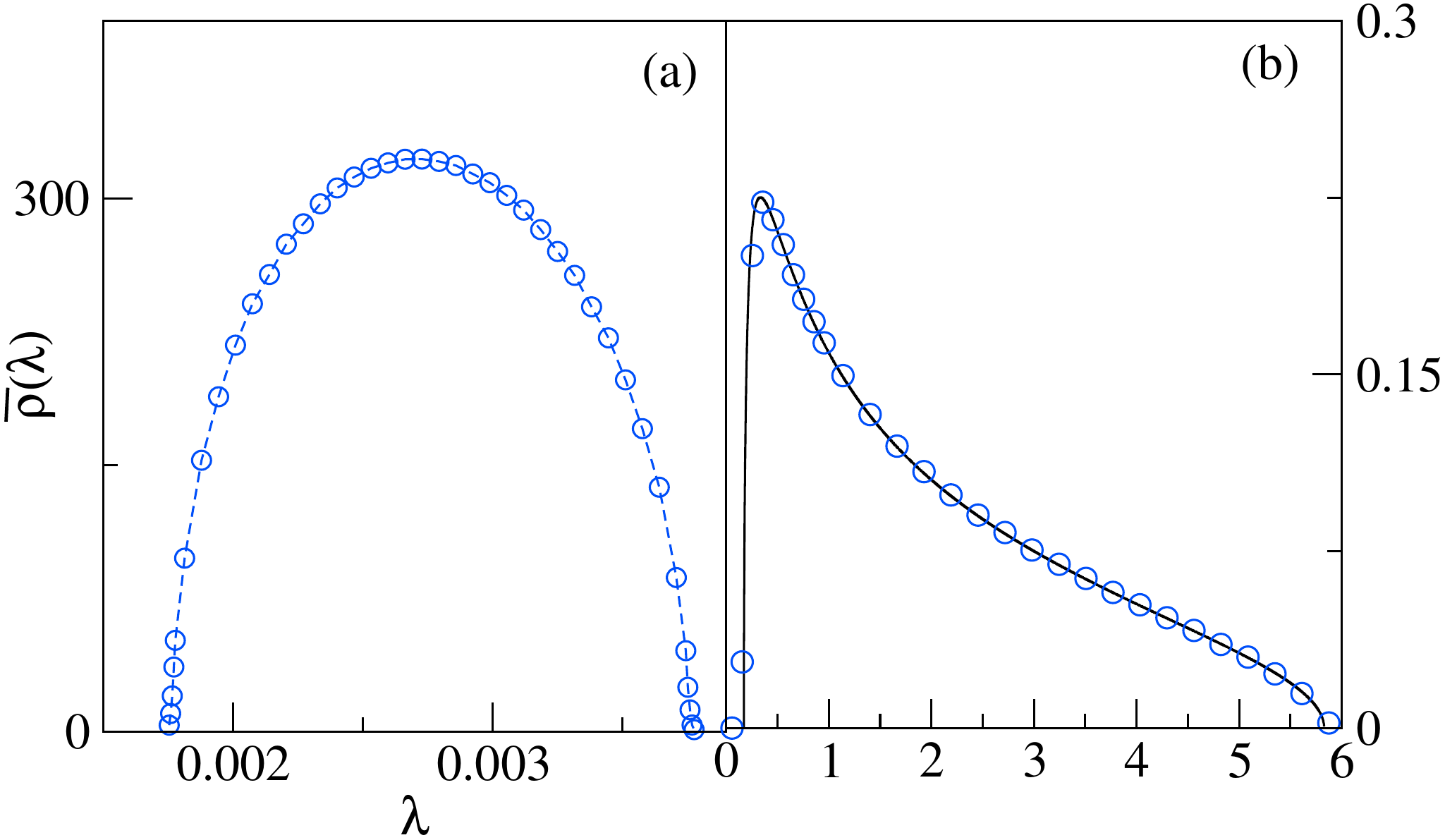}
                \caption{Density of eigenvalues of $\mathsf{C}^{(q)}$ where $q=1.001$, $N=1024$ and $\kappa=1/2$ for the WOE case. In (a) we show the density of emerging spectrum while in (b) we show the density of the former non-zero eigenvalues which is closely described by the Mar\'{c}enko Pastur law (\ref{denmp}) shown by a solid line. Both densities are shown on different scales. The density in Fig. \ref{dencq}(a) is normalized to $1-\kappa$ while the density in Fig. \ref{dencq}(b) is normalized to $\kappa$. Note that the density in Fig. \ref{dencq}(a) is not quite symmetric.}
\label{dencq}
\end{figure}

We illustrate some of these remarks with numerics for the WOE case. Consider a $1024\times 512$ random matrix $\mathsf{A}$ where the matrix elements are independent Gaussian variables with mean zero and variance one. Let the exponent of the power map (\ref{pmap}) be close to one, say, $q=1.001$. In Fig. \ref{dencq}(a) we show the density of the emerging spectra and in Fig. \ref{dencq}(b) we show the density of the eigenvalues near the Mar\'{c}enko-Pastur density which actually turns out to be very close to the latter.

For WOE, estimates for the first two moments of the emerging spectra have been derived in Ref. \cite{VRT}. We have also been able to extend our method to the equal-cross-correlation matrix model of CWOE. As we wish to make an expansion around $q=1$ we introduce the small parameter  $\alpha=(q-1)$ and define $\mathsf{C}^{(\alpha)}\equiv\mathsf{C}^{(q)}$ as defined in Eq. (\ref{pmap}). For small $\alpha$, $\mathsf{C}^{(\alpha)}$ may be expanded as
\begin{eqnarray}\label{LRC}
C_{jk}^{(\alpha)}&=&C^{(0)}_{jk}\exp\left[\frac{\alpha}{2}\text{ln}[(C^{(0)}_{jk})^{2}\right]
\nonumber\\
&=& C_{jk}+\frac{\alpha}{2}\,C_{jk}\,\text{ln}(C_{jk}^{2})\left[1+\mathcal{O}\left(\alpha\right)\right],
\end{eqnarray}
where in the second equality we drop the superscript, using from now on $\mathsf{C}$ for $\mathsf{C}^{(0)}$. Next, we expand the eigenvalues $\lambda_{j}(\alpha)$, of $\mathsf{C}^{(\alpha)}$, as
\begin{equation}\label{asigvl}
\lambda_{j}(\alpha)=\lambda_{j}(0)+\alpha\,(\delta\lambda_{j})[1+\mathcal{O}(\alpha)].
\end{equation}
Here the $\lambda_{j}(0)'$s are the eigenvalues of $\mathsf{C}$, for $j=1,..., N$ and
the $\alpha(\delta\lambda_{j})'$s are the leading order corrections coming from the power
map. For a short time horizon, i.e. $T<N$, $\lambda_{j}(0)=0$ for $j\le N-T$. For small
$\alpha$, we assume that the statistics of the relative changes in the eigenvalues is
dominated by the linear term. Bearing this in mind, we derive estimates for the moments
of the $\alpha(\delta\lambda_{j})'$s in the linear response regime. We refer to
$\alpha(\delta\lambda_{j})$ as the eigenvalues of the emerging spectrum, for $j\le N-T$,
otherwise as corrections of the former non-zero eigenvalues or non-zero eigenvalue corrections.
We consider $\alpha> 0$.

\begin{figure}
        \centering
               \includegraphics [width=0.9\textwidth]{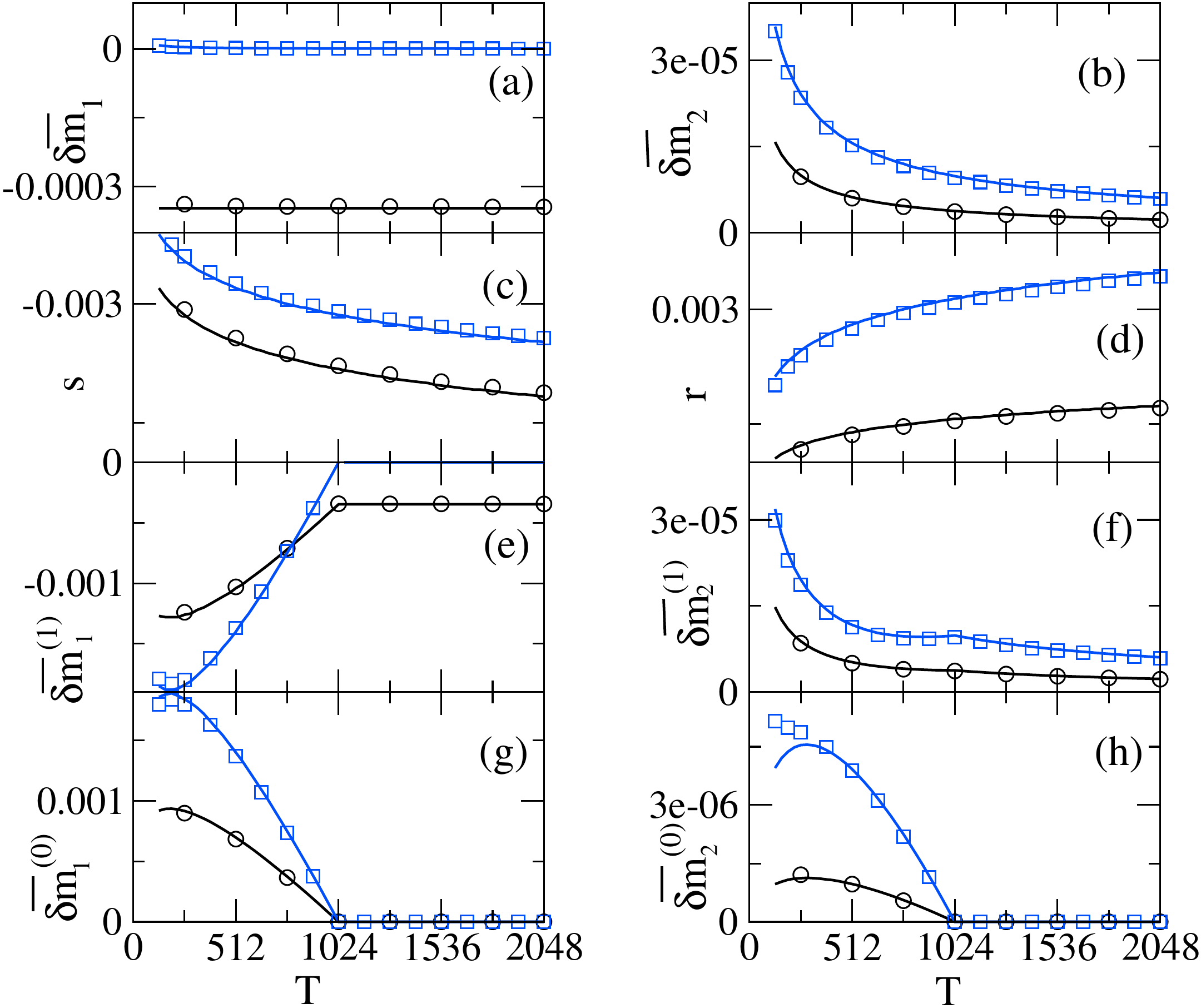}
                \caption{Comparison of theoretical and numerical results for CWOE where the non-random matrix elements are $\xi_{jk}=\delta_{jk}+c\,(1-\delta_{jk})$ and for WOE. We chose $c=0.5$ and varied $T$ for fixed $N=1024$. In this figure we compare the moments only for the bulk densities. CWOE numerics are shown by circles and WOE numerics are shown by blue squares. In (a) and (b) we compare the first two moments $\overline{\delta m_{1}}$ and $\overline{\delta m_{2}}$ obtained from the numerical simulations with the theory (\ref{m1m2ap}). In (c) and (d) we compare numerical results of scaling and shift-parameters $s$ and $r$. In (e) and (f) we compare numerical results of $\overline{\delta m^{\text{mp}}_{1}}$ and $\overline{\delta m^{\text{mp}}_{2}}$ with theory (\ref{m11m12ap}). Finally, in (g) and (h) we compare $\overline{\delta m^{(0)}_{1}}$ and $\overline{\delta m^{(0)}_{2}}$ with the theory (\ref{m01m02ap}). We refer to \cite{VRT} for the details of CWOE theoretical results.}
\label{m1m2}
\end{figure}

For {all} the  eigenvalue corrections, we define the moments as
\begin{eqnarray}\label{delm}
\overline{\delta m_{n}}&\equiv&\frac{\alpha^{n}}{N}\overline{\sum_{j=1}^{N}\,(\delta\lambda_{j})^{n}}.
\end{eqnarray}
Note that in the linear response regime $\overline{\delta m_{n}}$ may also be estimated by
\begin{eqnarray}\label{avCq}
\overline{\delta m_{n}}
&\simeq&\text{$\alpha^{n}$ term in}
\frac{1}{N}\left[\overline{\text{Tr}\,\left(\mathsf{C}^{(\alpha)}\right)^{n}}-\overline{\text{Tr}\,\left(\mathsf{C}\right)^{n}}\right].
\end{eqnarray}
Defining $\overline{\delta m^{(0)}_{n}}$ as the moments of emerging spectra and $\overline{\delta m^{(1)}_{n}}$ as the moments of non-zero eigenvalue corrections, we write
\begin{eqnarray}\label{m0pm1p}
\overline{\delta m}_{n}=
\overline{\delta m^{(0)}_{n}}+\overline{\delta m^{(1)}_{n}}.
\end{eqnarray}
{An important, though almost trivial remark is that $\overline{\delta m^{(1)}_{n}}\equiv\overline{\delta m_{n}}$ for $T\ge N$.} In linear response regime we obtain \cite{VRT}
\begin{eqnarray}\label{m1m2ap}
\overline{\delta m_{1}}&\sim&\frac{\alpha}{T},
\nonumber\\
\overline{\delta m_{2}}&\sim&\frac{\alpha^{2}}{4\kappa}\left([\text{log}(T)+c_{1}]^{2}+c_{2}\right),
\end{eqnarray}
where $c_{1}=\gamma+\text{log}(2)-2= -0.729637...$ and $c_{2}=\pi^{2}/2-4=0.934802...$, for $\gamma$ being the Euler constant, and
\begin{eqnarray}\label{m11m12ap}
\overline{\delta m^{(1)}_{1}}&=&\kappa\overline{\delta m_{1}}+s(1-\kappa),
\nonumber\\
\label{delm22}
\overline{\delta m^{(1)}_{2}}&=&\kappa\overline{\delta m_{2}}-\kappa\overline{\delta m_{1}}^{2}+\frac{\overline{(\delta m^{(1)}_{1})}^{2}}{\kappa}.
\end{eqnarray}
Here $s\sim -\frac{\alpha}{2}\sqrt{[\text{log}(T)+c_{1}]^{2}+c_{2}}$ is the scaling-parameter. In the derivation we have also used a shifting-parameter, $r=\overline{\delta m}_{1}-s(1-c)$. Using (\ref{m0pm1p}), estimation of the moments of the emerging spectra becomes trivial. For large $T$
and $\kappa\le1$, we get
\begin{eqnarray}\label{m01m02ap}
\overline{\delta m^{(0)}_{1}}&=&-s(1-\kappa),
\nonumber\\
\overline{\delta m^{(0)}_{2}}&=&s^{2}(1-\kappa).
\end{eqnarray}
Note, that for small values of $\kappa$, the error in our approach becomes large and linear response theory fails.

For equal cross-correlation matrix model for the CWOE, $\xi$ is a dense matrix and its eigenvalues, the $\xi_{j}$'s, are simply given by $\xi_{j}=(1-c)$ for $1\le j\le N-1$ and $\xi_{N}=Nc+1-c$.
For this spectrum, Eqs. (\ref{ResMp}, \ref{denCWE}) yields the density \cite{vp2010}
\begin{eqnarray}
\label{eqdeneqc}
\overline{\rho}(\lambda)&=&\overline{\rho'}(\lambda)+\delta\left(\lambda-\frac{(Nc+1-c)(Nc\kappa+1-c)}{Nc\kappa}\right),
\nonumber\\
\overline{\rho'}(\lambda)&=&
\kappa\frac{\sqrt{(\lambda_{+}-\lambda)(\lambda-\lambda_{-})}}{2\pi (1-c)\lambda},
\end{eqnarray}
where $\lambda_{\pm}=(1-c)(\kappa^{-1/2}\pm 1)^2$. {The similarity of $\overline{\rho'}(\lambda)$
with the Mar\'{c}enko-Pastur law (\ref{denmp}) is evident here, with the only difference of a
factor of $(1-c)$ in the place of $\sigma^{2}$ in (\ref{denmp})}. Note here that the bulk
density changes with the correlations, which is different from what we have seen for the non-symmetric correlation matrices in the previous section. The delta function appears in the above result as long
as $c\ge (N\sqrt{\kappa})^{-1}$. Using these we find
\begin{equation}
\overline{\delta m^{(1)}_{1}}=\kappa \overline{\delta m_{1}}+(1-c)s(1-\kappa),
\end{equation}
while the result for the second moment is the same as obtained in Eq. (\ref{delm22}).                                                                               

\section{Outlook}\label{outl}
The analysis of singular correlation matrices surfaces as an essential element in order to apply correlation analysis to quasi-stationary situations. These are not only the typical ones in finance, but also in a wide variety of fields displaying complex dynamics with state transitions. While the concept of states of a market is fairly new \cite{Thomas2012}, such situations are quite common in dynamical systems \cite{crossHrev} and play an important role in chemical reactors \cite{Arik1, Arik2}. Many other situations come to mind readily, such as social evolution, biological systems, climate, seismology and vulcanology to mention only a few. In all these cases obviously great importance must be attached to the identification of shorter transients and or precursors of transitions that in some cases may be quite dramatic. Market meltdowns, explosions in chemical reactors, major volcanic eruptions or social upheavals are examples of such dramatic events, but even an abrupt change of the output products in a chemical reactors may be quite expensive. It must be noted that time-lagged correlations is also of great interest in this context but less on this subject is known \cite{vinasm}. The problem is that we will get a matrix for each time delay and in principle nothing hinders us to consider different time delays for different time series. Yet that obviously is opening Pandoras box. We would need compelling dynamical reasons to take such a step, while the present analysis really is directed to an unbiased analysis, i.e., toward previous ignorance of the dynamics.


\begin{theacknowledgments}
 The authors acknowledge financial support by PAPIIT/DGAPA/UNAM under project IG101113 and one of us (Vinayak) is grateful for a postdoctoral fellowship provided by DGAPA/UNAM.
\end{theacknowledgments}

\section{Appendix}
\subsection{Derivation of the Pastur Density for CWOE}\label{Pastur}
To keep notation handy here we will be a little inconsistent in this respect. We use angular-brackets for the spectral averaging. For instance, we use $\langle\{.\}\rangle_{K}=K^{-1}\text{tr}\{.\}$. We prefer to
use over line if deal with a more general quantity such as a resolvent $\overline{G}_{L}(z)$:
\begin{equation}
\overline{G}_{L}(z)=\llan L\overline{(z-\mathsf{C})^{-1}}
\rran.
\end{equation}
Here $L$ is an arbitrary nonrandom matrix. We note that for large $z$, the resolvent may be expressed in terms of moments, $\overline{\mathbf{m}^{L}_{p}}$, as
\begin{eqnarray}\label{Resmom}
\overline{G}_{L}(z)&=&\sum_{p=0}^{\infty}\dfrac{\llan\overline{ L\mathbf{C}^{p}}\rran_{N}}{z^{p+1}}
\nonumber\\
&=& \sum_{p=0}^{\infty}\dfrac{\overline{\mathbf{m}^{L}_{p}}}{z^{p+1}}.
\end{eqnarray}
Using the JPD (\ref{jpdajk}) we derive the following exact results, valid for arbitrary fixed matrices
$\Phi$ and $\Psi$,
\begin{eqnarray}\label{Iden1}
\dfrac{1}{T} 
\lan\overline{ \mathsf{B} \Phi \mathsf{B}^{t}\Psi}
\ran_{N}
&=&\sigma^{2}\llan \Phi \rran_{T} \llan \Psi \rran_{N},
\\
\label{Iden2}
\lan\overline{ \mathsf{B} \Phi \mathsf{B}\Psi}
\ran_{N}
&=&\sigma^{2}\llan \Phi^{t}\Psi\rran_{N},
\\
\label{Iden3}
\overline{\llan \mathsf{B} \Phi\rran \llan \Psi \mathsf{B}^{t}
\rran}_{N}
&=&\dfrac{\sigma^{2}}{N}
\llan \Psi \Phi \rran_{N},
\\
\label{Iden4}
\overline{\llan \mathsf{B} \Phi\rran \llan \mathsf{B} \Psi
\rran}_{N}
&=&\dfrac{\sigma^{2}}{N}
\llan \Psi^{t} \Phi \rran_{N}.
\end{eqnarray}
Here dimensions of $\Phi$ and $\Psi$ are suitably adjusted in identities (\ref{Iden1}-\ref{Iden4}).

It is trivial to obtain the first two terms in the expansion (\ref{Resmom});
$\overline{\mathbf{m}^{L}_{0}}=1$ and by definition
$\overline{\mathbf{m}^{L}_{1}}=\sigma^{2}\llan L\xi\rran$. In what follows, for
$p\ge2$, we consider only the leading order terms resulting from ensemble averaging,
dropping terms of order $\mathcal{O}(N^{-1})$. For instance, it follows from
Eqs. (\ref{Iden1},\ref{Iden2}) that only the binary associations of $\mathsf{B}$
with $\mathsf{B}^{t}$ produce leading order terms. Therefore, using the identity
(\ref{Iden1}), we calculate
\begin{eqnarray}\label{mompge2}
\overline{\mathbf{m}^{L}_{p}}&=&\llan \overline{L\xi^{1/2}\underbracket[1pt]{\mathsf{BB}^{t}}\xi^{1/2}\mathsf{C^{p-1}}}\rran
\nonumber\\
&+&
\sum_{n=0}^{p-2}
\llan\overline{L\xi^{1/2}\underbracket[1pt]{\mathsf{BB}^{t}\xi^{1/2}\mathsf{C^{n}} \xi^{1/2}\mathsf{BB}^{t}} \xi^{1/2}\mathsf{C^{p-n-2}}}\rran
\nonumber\\
&=&
\sigma^{2}\overline{\mathbf{m}^{L\xi}_{p-1}}+
\dfrac{\sigma^{2}}{\kappa} \sum_{n=0}^{p-2}\overline{\mathbf{m}}_{n+1}\,\overline{\mathbf{m}^{L\xi}_{p-n-2}},
\end{eqnarray}
where under-brackets are used to describe the binary associations yielding leading order terms.
Note that the equality in the last equation is valid in leading order and terms resulting
from the binary associations across the traces in the intermediate steps have been ignored;
see (\ref{Iden3},\ref{Iden4}). Using Eq. (\ref{mompge2}) in Eq. (\ref{Resmom}) and after
rearranging series, we find
\begin{eqnarray}
z\overline{G}_{L}(z)&=&\llan\,L\,\rran+\dfrac{\sigma^{2}}{\kappa}\left(\kappa-1+z\overline{G}(z)\right)\overline{G}_{L\xi}(z).
\end{eqnarray}
Substituting $\L= L\,\left(z-\dfrac{\sigma^{2}}{\kappa}\left(\kappa-1+z\overline{G}(z)\right)\xi\right)^{-1}$, we get a self-consistent equation:
\begin{eqnarray}
\nonumber\\
\overline{G}_{\L}(z)
&=&
\llan \L\,\left(z-\dfrac{\sigma^{2}}{\kappa}\left(\kappa-1+z\overline{G}(z)\right)\xi\right)^{-1}
\rran.
\end{eqnarray}
Finally, for $\L=\mathbf{1}$ we get the Pastur self-consistent equation or the Pastur density (\ref{ResMp}).

\subsection{Details of the portfolio optimization study} \label{app:opt}
The weight vector of the minimal variance portfolio is calculated as
\begin{equation}
w_\mathrm{opt}=\frac{\Sigma^{-1} e}{e^t \Sigma^{-1} e} \ ,
\label{eq:wopt}
\end{equation}
where $e$ is a vector of length $N$ with all entries set to one, $e^t$ denotes
the transposed vector. In order to calculate the optimal weights (\ref{eq:wopt}),
we need to know the covariance matrix $\Sigma$ of the $N$ stock returns.
In practice, this covariance matrix has to be estimated using historical data.
Here we consider a model setting with $N=100$ stocks, a model correlation matrix
$C_0$ with 5 blocks of size 20, corresponding to industry sectors and randomly
distributed but fixed standard deviations $\sigma_k$.
In a factor model that reflects our correlation matrix $C_0$, we simulate time
series of length $N$, estimate the correlation matrix $C^\mathrm{(samp)}$ and
apply the power map to arrive at the matrix $C^{(q)}$ with entries
\begin{equation}
C^{(q)}_{kl}={\rm sign}\,C^\mathrm{(samp)}_{kl} \, \big| C^\mathrm{(samp)}_{kl} \big|^q \ .
\end{equation}
We multiply with the standard deviations to get the elements of the covariance matrix,
\begin{equation}
\widehat{\Sigma}_{kl}=\sigma_k \sigma_l C^{(q)}_{kl} \ .
\end{equation}
With this covariance matrix we calculate the weights
\begin{equation}
\widehat{w}_\mathrm{opt}=\frac{\widehat{\Sigma}^{-1} e}{e^t \widehat{\Sigma}^{-1} e} \ .
\label{eq:wopthat}
\end{equation}
Using the model correlation matrix $C_0$ and the corresponding covariance matrix
$\Sigma_0=\sigma C_0 \sigma$, with $\sigma={\rm diag}(\sigma_1, \ldots, \sigma_N)$,
the actual portfolio variance for the weights (\ref{eq:wopthat}) reads
\begin{equation}
\Omega^2=\widehat{w}_\mathrm{opt}^{\,t} \Sigma_0 \widehat{w}_\mathrm{opt} \ ,
\end{equation}
whereas the minimal variance $\Omega_0^2$ is obtained by calculating the optimal
weights (\ref{eq:wopt}) for $\Sigma_0$.


\begin{thebibliography}{99}
\bibitem{bulk} It is understood that in the case of separation of $n<<N$ eigenvalues the bulk is normalized to $1-n/N$.
\bibitem{sing} Singular values of $\mathsf{AB}^{t}$ are the positive square root of the nonzero eigenvalues of $\mathbf{C}$.
\bibitem{dbulk} Density of the separated eigenvalues for CWOE is also Gaussian.
\end{thebibliography}
\end{document}